\documentclass[journal]{IEEEtran}
\usepackage{color}
\usepackage{float}
\usepackage{tabularx,colortbl}
\usepackage{setspace}
\usepackage{cite,graphicx,amsmath,psfrag,bm}
\usepackage{subfigure}
\usepackage[usenames,dvipsnames]{xcolor}
\usepackage{mathabx}
\usepackage{cancel}
\usepackage{epstopdf}
\usepackage{amsmath}

\usepackage[process=auto]{pstool}
\usepackage[normalem]{ulem}
\graphicspath{{Figs/}}

\newcommand{\figref}{Fig.~\ref}

\ifCLASSINFOpdf
\else
\fi

\begin{document}

\title{Flexible-Resolution, Arbitrary-Input and Tunable Rotman Lens Spectrum Decomposer~(RL-SD)}

\author{Xiaoyi~Wang,~\IEEEmembership{Student Member,~IEEE,}
        Alireza~Akbarzadeh,
        Lianfeng~Zou,~\IEEEmembership{Student Member,~IEEE,}
        and~Christophe~Caloz,~\IEEEmembership{Fellow,~IEEE}% <-this % stops a space
\thanks{
The authors are with the Department
of Electrical Engineering, Polytechnique Montr\'{e}al,
Montr\'{e}al, Qu\'{e}bec, Canada.(e-mail:xiaoyi.wang@polymtl.ca).}% <-this % stops a space
}

\maketitle

\begin{abstract}
We present an enhanced design -- in terms of resolution flexibility, input port position arbitrariness and frequency-range tunability -- of the planar Rotman lens spectrum decomposer (RL-SD). This enhancement is achieved by manipulating the output port locations through proper sampling of the frequency-position law of the RL-SD, inserting a calibration array compensating for frequency deviation induced by input modification and introducing port switching, respectively. A complete design procedure is provided and two enhanced RL-SD prototypes, with uniform port distribution and uniform frequency resolution, respectively, are numerically and experimentally demonstrated.
\end{abstract}


\IEEEpeerreviewmaketitle

\section{Introduction}\label{sec:intro}

Spectral Decomposition (SD) is a fundamental optical process according to which white light is spatially  split into its constituent frequencies. It is widely found in natural phenomena, such as in rainbows, which are caused by the interplay of reflection, refraction and dispersion of light in water droplets~\cite{Book:Waldman_2002_LightIntro}, or in soap bubbles and gas spills, which are caused by interferences due thin film thickness gradients~\cite{Book:Knittl_1976_ThinFilms}. It is also produced by human-made devices, such as prisms and diffraction gratings~\cite{Book:Born_2013_PrinciplesofOptics,Book:Saleh_2007_FundamentalsPhotonics}, and may be achieved in any part of the electromagnetic spectrum. SD devices are used in various applications, including  colorimetry~\cite{Jour:Kok_1971_Colorimeter}, real-time spectrum analysis~\cite{JOUR:2009_Gupta_TMTT_RTSA}, laser wavelength tuning~\cite{JOUR:Hard_1970_laser,JOUR:Hansch_1972_Laser_tuning}, filtering~\cite{Jour:White_1947_Filter_Gratings,Jour:Knop_1978_Color_Filtering}, wavelength division multiplexing~\cite{JOUR:Brackett_1990_WDM,JOUR:1978_Henry_TMTT_Grating_Multiplexer}, and antenna array beam forming~\cite{Jour:Zmuda_1997_beamformer}.

Most of the conventional SD devices are based on prisms and diffraction gratings, which are three-dimensional components that are bulky, expensive and incompatible with integrated circuit technology. The availability of two-dimensional, or \emph{planar}, implementations would resolve these issues and hence dramatically widen the range of applications of SD. Several efforts have been dedicated to realize planar SD devices, particulary using lumped-element structures~\cite{JOUR:2009_Momeni_TMTT_Electrical_Prism, JOUR:2008_Afshari_TCS_2D_LC_Lattice} and leaky-wave antennas~\cite{JOUR:2009_Gupta_TMTT_RTSA,JOUR:2015_Gomez-Tornero_TMTT_SIW_Multiplexer}. The most practical attempt has probably been the SD reported by Zhang and Fusco~\cite{Conf:Fusco_Multiplerxer_2012}, which transforms the beam forming operation of a Rotman lens (RL) into a SD operation using a reflecting transmission line array.

In~\cite{Conf:Fusco_Multiplerxer_2012}, the emphasis was on generating orthogonal spectrally decomposed waveforms. In this paper, we enhance the SD capability of the Zhang-Fusco RL-SD in terms of resolution flexibility, input port position arbitrariness and frequency-range tunability, by manipulating the output port locations, inserting a calibration array and introducing port switching, respectively. With such enhancements, the RL-SD becomes a promising device for integrated systems involving real-time spectrum analysis.

\section{Proposed Enhanced RL-SD}\label{sec:Concept}
\subsection{Recall of Rotman Lens (RL)}\label{sec:RL}

\begin{figure}[h!t]
    \centering
    \includegraphics[width=0.9\columnwidth]{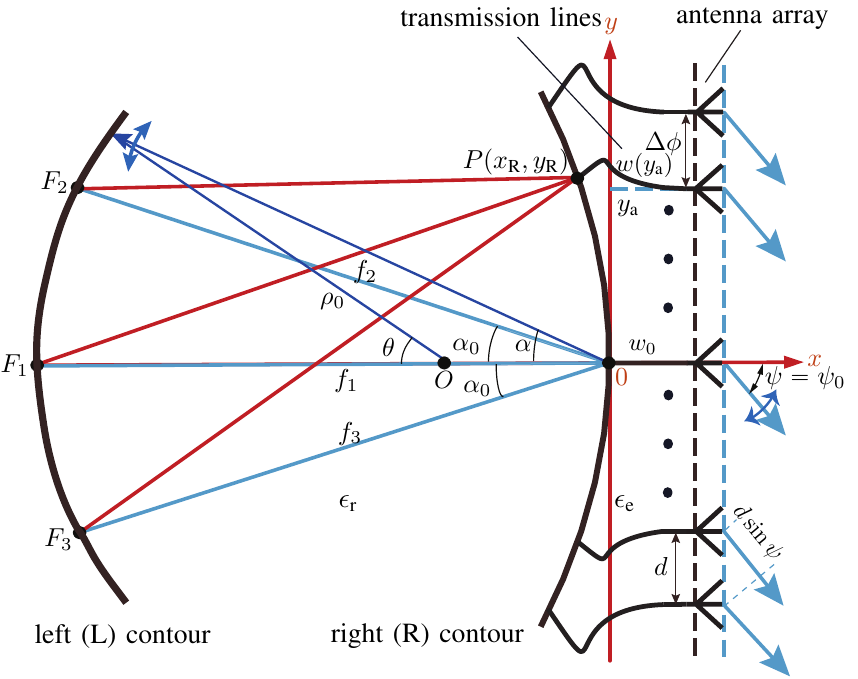}
         %\psfragfig[width=0.98\linewidth, trim={-0.1in 0in -0.4in 0in}]{Figs/RotmanLens}{
%         \psfrag{A}[c][c][0.75]{$F_1$}
%         \psfrag{B}[c][c][0.75]{$F_2$}
%         \psfrag{C}[c][c][0.75]{$F_3$}
%         \psfrag{P}[c][c][0.75]{$P(x_\text{R},y_\text{R})$}
%         \psfrag{x}[c][c][0.75]{\color{Bittersweet}{$x$}}
%         \psfrag{y}[c][c][0.75]{\color{Bittersweet}{$y$}}
%         \psfrag{O}[c][c][0.75]{\color{Bittersweet}{$0$}}
%         \psfrag{L}[c][c][0.8]{left (L) contour}
%         \psfrag{S}[r][r][0.8]{right (R) contour}
%         \psfrag{D}[l][l][0.75]{$\psi=\psi_0$}
%         \psfrag{R}[r][r][0.75]{$\alpha_0$}
%         \psfrag{G}[r][r][0.75]{$\theta$}
%         \psfrag{m}[r][r][0.75]{$f_1$}
%         \psfrag{n}[r][r][0.75]{$f_2$}
%         \psfrag{q}[r][r][0.75]{$\rho_0$}
%         \psfrag{Q}[c][c][0.72]{$O$}
%         \psfrag{T}[c][c][0.8]{transmission lines}
%         \psfrag{a}[c][c][0.8]{antenna array}
%         \psfrag{E}[l][l][0.75]{$y_\text{a}$}
%         \psfrag{W}[c][c][0.7]{$w(y_\text{a})$}
%         \psfrag{w}[c][c][0.7]{$w_0$}
%         \psfrag{r}[c][c][0.8]{$\epsilon_\text{r}$}
%         \psfrag{e}[c][c][0.8]{$\epsilon_\text{e}$}
%         \psfrag{k}[c][c][0.75]{$\alpha$}
%         \psfrag{t}[c][c][0.75]{$f_3$}
%         \psfrag{d}[c][c][0.75]{$d$}
%         \psfrag{V}[c][c][0.75]{$\Delta\phi$}
%         \psfrag{v}[c][c][0.68]{$d\sin{\psi}$}
%         \psfrag{v}[c][c][0.68]{$d\sin{\psi}$}
%        }
        \caption{Geometry and parameters of the Rotman lens (RL).}
   \label{FIG:RotmanLens}
\end{figure}

RLs are planar lenses that are generally used as beamforming feeding networks in antenna arrays. Given their combination of wide bandwidth, following from their true time delay\footnote{(See \figref{FIG:RotmanLens}) The Rotman lens is essentially a two-dimensional nondispersive structure, i.e. a propagation medium having a frequency-linear dispersion relation, $\beta_\text{r}(\omega)=\sqrt{\epsilon_\text{r}}k_0=\sqrt{\epsilon_\text{r}}\omega/c$, where $\epsilon_\text{r}$ is the lens permittivity. The antenna array radiation angle, assuming inter-element spacing $d$, is, from antenna array theory, $\psi=\sin^{-1}[\Delta\phi/(k_0d)]$, where $\Delta\phi=\beta_\text{r}(\omega)\Delta l+\beta_\text{e}(\omega)\Delta w$, with $\beta_\text{e}=\sqrt{\epsilon_\text{e}}k_0=\sqrt{\epsilon_\text{e}}\omega/c$ being the effective wavenumber of the transmission lines, and $\Delta l$ and $\Delta w$ denoting the lens and line paths differences to adjacent array elements. Then $\Delta\phi/k_0=\sqrt{\epsilon_\text{r}}\Delta l+\sqrt{\epsilon_\text{e}}\Delta w=\text{const.}$, so that $\psi$ is independent of frequency, which theoretically leads to infinite bandwidth operation. In a true \emph{phase} shifter system, in contrast, we would have a wavenumber function of the form $\beta_\text{r}(\omega)=ak_0+b=a\omega/c+b$, with $a$ and $b$ constant, leading to $\Delta\phi/k_0=a\Delta l+\sqrt{\epsilon_\text{e}}\Delta w+(b\Delta l c)\omega\neq\text{const.}$ and therefore a frequency-dependent or beam-squinting array.}
nature~\cite{Jour:Rotman_ProcIEEE_TrueTimeDelay}, and low-cost beam switching capability, compared to expensive continuous-scanning systems, they suit applications such as ultra-wide band communication systems, collision avoidance radars, and remote-piloted vehicles~\cite{JOUR:Rotman_Rotmanlens_1963,JOUR:Hansen_Rotmanlenses_1991,JOUR:2014_CJE_Vashist_ReviewRotamLens,Thesis:Dong_RotmanLens_2009}.

Figure~\ref{FIG:RotmanLens} shows the geometry and parameters of an RL. It is essentially a parallel-plate waveguide structure, with a substrate of permittivity $\epsilon_\text{r}$, delimited by two geometrical contours: a left (L) contour, supporting the beamformer input ports (not shown explicitly in this figure), and a right (R) contour, branching out to an array of transmission lines feeding the antenna array elements. The angle $\alpha$ locates the position of the input ports, and $w(y_\text{a})$ denotes the transmission line length at the output $y_\text{a}$, with $w_0=w(y_\text{a}=0)$. Excitation of the port located at an angle $\alpha$ leads the antenna array to radiate at the angle $\psi$, and switching between different ports provides radiation scanning.

The lens is designed by considering plane wave excitation from the right of the RL structure, for convenient application of geometrical optics. In this situation, $F_1$, $F_2$ and $F_3$ are three perfect focal points along the left contour, with respective focal lengths $f_1$, $f_2$ and $f_3=f_2$, and focal angles $0$, $\alpha_0$ and $-\alpha_0$, where $\alpha_0$ corresponds to $\psi_0$. Perfect focalization is achieved at $F_1$, $F_2$ and $F_3$ if the optical path from the incident wave (from the right, corresponding to inverted blue arrows in Fig.~\ref{FIG:RotmanLens}) phase front to each of these points is the same for all the points $P(x_\text{R},y_\text{R})$ connected to the antenna array on the right contour, particularly to the optical path from the center ($y=0$) antenna element, i.e.
\begin{subequations}
\label{EQ:RLBasics}
\begin{equation}
\label{EQ:RLBasic1}
F_1:w(y_\text{a})\sqrt{\epsilon_\text{e}}+PF_1\sqrt{\epsilon_\text{r}}=w_0\sqrt{\epsilon_\text{e}}+f_1\sqrt{\epsilon_\text{r}},\\
\end{equation}
\begin{equation}
\label{EQ:RLBasic2}
F_2:y_\text{a}\sin\psi_0+w(y_\text{a})\sqrt{\epsilon_\text{e}}+PF_2\sqrt{\epsilon_\text{r}}=w_0\sqrt{\epsilon_\text{e}}+f_2\sqrt{\epsilon_\text{r}},\\
\end{equation}
\begin{equation}
\label{EQ:RLBasic3}
F_3:-y_\text{a}\sin\psi_0+w(y_\text{a})\sqrt{\epsilon_\text{e}}+PF_3\sqrt{\epsilon_\text{r}}=w_0\sqrt{\epsilon_\text{e}}+f_3\sqrt{\epsilon_\text{r}},
\end{equation}
\end{subequations}
where $\epsilon_\text{e}$ is the effective permittivity of the transmission lines. This represents a system of three equations in three unknowns, $x_\text{R}$, $y_\text{R}$ and $w$, which is solved in the Appendix~\ref{sec:RL_Synth}, for the right contour $[x_\text{R}(y_\text{a}),y_\text{R}(y_\text{a})]$ and transmission line lengths $[w(y_\text{a})]$. The shape of the left contour is also given in Appendix~\ref{sec:RL_Synth}.

Let us finally define, for later use, the parameter
\begin{equation}\label{EQ:Gamma}
\gamma=\frac{\sin{\psi_0}}{\sin\alpha_0}\approx\frac{\sin{\psi}}{\sin\alpha},
\end{equation}
that relates the input port position angle ($\alpha$) and the beam steering angle ($\psi$)~\cite{JOUR:Rotman_Rotmanlens_1963}. In~\eqref{EQ:Gamma}, the approximation and equality account for the fact that the beamforming operation is generally approximate, while being perfect only at the pairs $(0,0)$ and $(\pm\alpha_0,\pm\psi_0)$, corresponding to the aforementioned perfect focal points $F_1$, $F_2$ and $F_3$, respectively.

\subsection{Basic RL-SD Description}\label{sec:RLSD_Descr}

\begin{figure}[h!t]
    \centering
    \includegraphics[width=0.9\columnwidth]{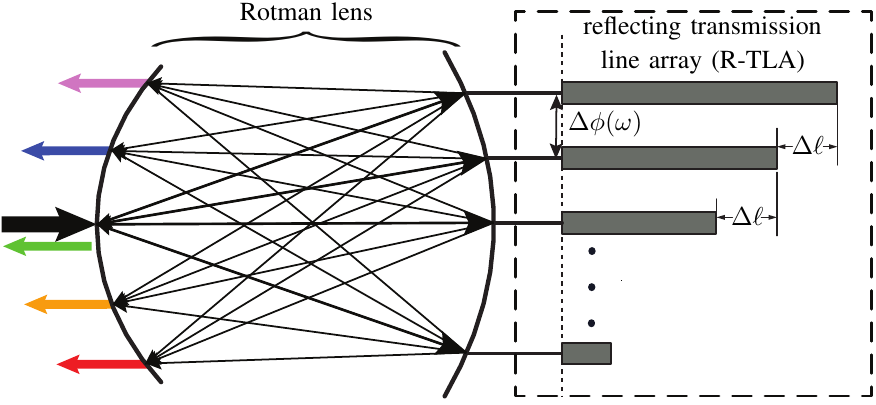}
         %\psfragfig[width=1\linewidth, trim={0in 0in 0in 0in}]{Figs/RotmanLensDecomposer}{
%         \psfrag{A}[c][c][0.75]{Rotman lens}
%          \psfrag{D}[c][c][0.75]{reflecting transmission}
%          \psfrag{G}[c][c][0.75]{line array (R-TLA)}
%          \psfrag{E}[c][c][0.75]{$\Delta\ell$}
%          \psfrag{F}[l][l][0.75]{$\Delta\phi(\omega)$}
%         }
        \caption{Rotman lens spectrum decomposition (RL-SD) reported in~\cite{Conf:Fusco_Multiplerxer_2012}. The connecting line segments between the lens and the RLTA represent here, and in all forthcoming figures, ideal connections with zero electrical length.}
   \label{FIG:FuscoRLSD}
\end{figure}

The Zhang-Fusco RL-SD~\cite{Conf:Fusco_Multiplerxer_2012} is depicted in~\figref{FIG:FuscoRLSD}. In this device, the antenna array in \figref{FIG:RotmanLens} has been replaced by an open-ended reflecting transmission line array (R-TLA), and the output ports with decomposed frequencies are on the left contour of the RL, as the input port, which is kept fixed for a specific spectral analysis.

Assuming that the length difference between adjacent transmission lines is $\Delta\ell$, as shown in~\figref{FIG:FuscoRLSD}, the corresponding phase difference after reflection from the R-TLA is
\begin{equation}\label{DeltaPhi}
\Delta\phi(\omega)=2\beta_\text{e}(\omega)\Delta\ell,
\end{equation}
where the factor 2 accounts for the reflection round trip and $\beta_\text{e}(\omega)$ is the effective wavenumber of the transmission lines,
\begin{equation}\label{EQ:BetaE}
\beta_\text{e}(\omega)
=k_0\sqrt{\epsilon_\text{e}}
=\frac{\omega}{c}\sqrt{\epsilon_\text{e}}.
\end{equation}
Calling the frequency and wavelength of the output central port (green arrow in \figref{FIG:FuscoRLSD}) $\omega_0$ and $\lambda_0$, respectively, and assuming that
\begin{equation}\label{DeltaL}
\Delta\ell = N\frac{\lambda_0}{2}=N\frac{\pi c}{\omega_0},\quad N=1,2,3,...
\end{equation}
leads, upon substitution of~\eqref{EQ:BetaE} and~\eqref{DeltaL} into~\eqref{DeltaPhi}, to the frequency-dependent phase shift response
\begin{equation}\label{DeltaPhi1}
\Delta\phi(\omega)=2\beta_\text{e}(\omega)\Delta\ell=2\pi N\frac{\omega}{\omega_0}.
\end{equation}

The overall operation of the RL-SD may now be understood as follows. The signal to be spectrally analyzed (represented by the black arrow in \figref{FIG:FuscoRLSD}) is injected at the central port of the device. The corresponding wave propagates towards the right across the lens and hence splits to reach all the points of the right contour of the lens. At this stage, each split wave includes the entire spectrum of the input signal. Next, these waves are reflected by the R-TLA, which spectrally decomposes them according to the phase shift function~\eqref{DeltaPhi1}. From~\eqref{DeltaPhi1}, the frequency $\omega=\omega_0$ of the input signal corresponds to the phase gradient $\Delta\phi(\omega_0)=2\pi N$, and the corresponding reflected waves will therefore constructively interfere at the central focal point, $F_1$, so that the part of the signal spectrum with $\omega=\omega_0$ emerges at that port (green arrow). The parts of the signal spectrum with $\omega>\omega_0$ and $\omega<\omega_0$ essentially correspond to phase gradients $0<\Delta\phi(\omega)<\pi$ and $-\pi<\Delta\phi(\omega)<0$, respectively, hence leading to spectral decomposition to the upper and lower ports of the device, respectively, as illustrated in \figref{FIG:FuscoRLSD}.

\subsection{Resolution and Input Port Diversification}
\label{sec:resol_in_port_divers}

As mentioned in Sec.~\ref{sec:intro} and as will be shown in Sec.~\ref{sec:synthesis}, the RL-SD described in Sec.~\ref{sec:RLSD_Descr} suffers from rigid resolution and inconvenient input port location, which are the two main issues addressed by the paper.

\begin{figure}[h!t]
    \centering
    \includegraphics[width=0.9\columnwidth]{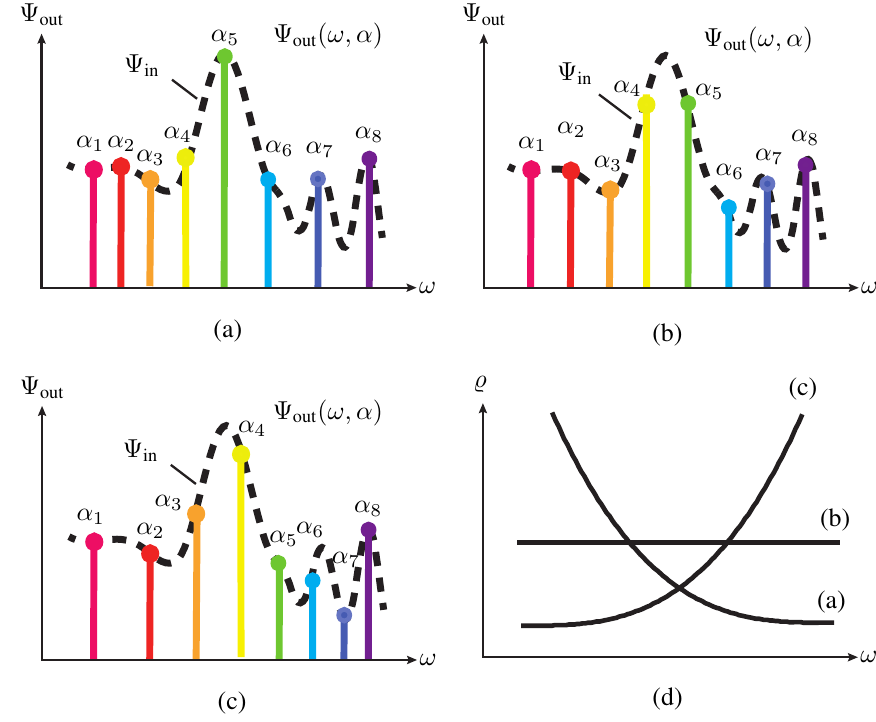}
        %\psfragfig[width=1\linewidth, trim={-0.2in 0in 0in 0in}]{Figs/SpectrumAnalysis}{
%        \psfrag{A}[c][c][0.75]{(a)}
%        \psfrag{B}[c][c][0.75]{(b)}
%        \psfrag{C}[c][c][0.75]{(c)}
%        \psfrag{D}[c][c][0.75]{(d)}
%        \psfrag{E}[c][c][0.75]{$\Psi_\text{in}$}
%        \psfrag{F}[c][c][0.75]{$\Psi_\text{out}(\omega,\alpha)$}
%        \psfrag{G}[l][l][0.75]{(c)}
%        \psfrag{H}[l][l][0.75]{(b)}
%        \psfrag{I}[l][l][0.75]{(a)}
%        \psfrag{J}[c][c][0.75]{$\varrho$}
%        \psfrag{j}[c][c][0.72]{$\Psi_\text{in}$}
%        \psfrag{u}[c][c][0.75]{$\alpha_1$}
%        \psfrag{v}[c][c][0.75]{$\alpha_2$}
%        \psfrag{w}[c][c][0.75]{$\alpha_3$}
%        \psfrag{x}[c][c][0.75]{$\alpha_4$}
%        \psfrag{y}[c][c][0.75]{$\alpha_5$}
%        \psfrag{a}[c][c][0.75]{$\alpha_6$}
%        \psfrag{b}[c][c][0.75]{$\alpha_7$}
%        \psfrag{z}[c][c][0.75]{$\alpha_8$}
%        \psfrag{Z}[c][c][0.75]{$\Psi_\text{out}$}
%        \psfrag{i}[c][c][0.75]{$\omega$}
%        }
        \caption{Spectrum analysis performed by different RL-SDs. (a)~Conventional RL-SD (\figref{FIG:FuscoRLSD}) with uniform port distribution ($\Delta\alpha=\text{const.}$), leading to decreasing resolution ($\varrho=1/\Delta\omega$) with increasing frequency ($\partial\varrho/\partial\omega<0$). (b)~RL-SD with nonuniform port distribution designed for uniform resolution ($\partial\varrho/\partial\omega=0$ or $\Delta\omega=\text{const.}$). (c)~RL-SD with nonuniform port distribution designed for reversed resolution compared to (b), i.e. increasing resolution with increasing frequency ($\partial\varrho/\partial\omega>0$) (d)~Resolution ($\varrho$) versus frequency ($\omega$) for the RL-SDs (a), (b) and (c).}
   \label{FIG:SpectrumAnalysis}
\end{figure}

Enhancing the resolution of the RL-SD will essentially consist in properly designing its frequency resolution function, $\varrho(\omega)$. Figure~\ref{FIG:SpectrumAnalysis} shows the type of spectrum analysis performed by both the RL-SD of \figref{FIG:FuscoRLSD} and proposed enhanced versions of it to an arbitrary signal with spectrum $\Psi_\text{in}(\omega)$. Given the discrete nature of the ports, the spectral analysis will be a discrete version or, more precisely, a quantized version of $\Psi_\text{in}(\omega)$, and the resolution function is then
\begin{equation}\label{EQ:Rho}
\varrho(\omega)=\frac{1}{\Delta\omega(\omega)},
\end{equation}
where $\Delta\omega(\omega)$ represents the frequency varying distance between adjacent sampling points. This section only describes the general idea, to highlight the contribution of the paper, while the exact analysis will be provided in Sec.~\ref{sec:synthesis}.

Figure~\ref{FIG:SpectrumAnalysis}(a) shows the response of an RL-SD with uniform angular port distribution ($\Delta\alpha=\text{const.}$), as implicitly assumed in \figref{FIG:FuscoRLSD}, a resolution that decreases with increasing frequency, or $\partial\varrho/\partial\omega<0$, as shown in \figref{FIG:SpectrumAnalysis}(d)~\cite{Conf:Fusco_Multiplerxer_2012}. In the particular case of the signal $\Psi_\text{in}(\omega)$ in \figref{FIG:SpectrumAnalysis}, this leads to over-sampling and under-sampling the lower and higher parts, respectively, of the signal spectrum. Note that uniform angular resolution, or $\Delta\alpha=\text{const.}$, does not mean uniform arc distribution. Indeed, the arc length, measured from the $-x$ axis is $\chi=\rho_0\theta$, since the left contour has been chosen as a circular arc (Appendix~\ref{sec:RL_Synth}). According to~\eqref{BeamCoutourParamPhi}, we have then $\chi(\alpha)=\rho_0(\arcsin(\frac{1-\rho_0}{\rho_0}\sin{\alpha})+\alpha)$, which is obviously not linearly proportional to $\alpha$

Since the signal to analyze is generally unknown, no a priori assumption can be made on its spectrum, and therefore designing the device for uniform resolution would generally be the best strategy. As a uniform port distribution leads to nonuniform resolution [\figref{FIG:SpectrumAnalysis}(a)], we can deduce that the ports must be properly \emph{nonuniformly} distributed to provide uniform resolution, corresponding to $\partial\varrho/\partial\omega=0$, as shown in Figs.~\ref{FIG:SpectrumAnalysis}(b) and \ref{FIG:SpectrumAnalysis}(d).

In the case where there would be some a priori knowledge on the general shape of signal spectrum, as for instance if one would know that there is higher variations at the higher part of the spectrum as in \figref{FIG:SpectrumAnalysis}, then one could deliberately distort the resolution function so that $\partial\varrho/\partial\omega>0$, as shown in Figs.~\ref{FIG:SpectrumAnalysis}(c) and \ref{FIG:SpectrumAnalysis}(d), where the resolution function leads to the best frequency sampling for the given input signal.

\begin{figure}[h!t]
    \centering
     \includegraphics[width=0.9\columnwidth]{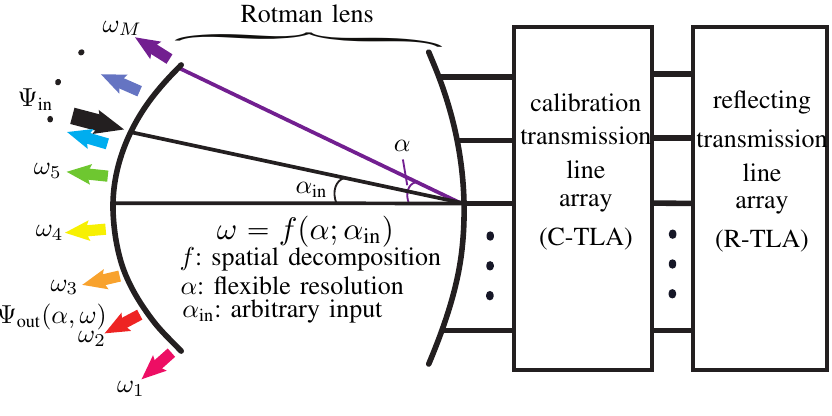}
       % \psfragfig[width=0.95\linewidth, trim={-0.2in 0in 0in 0in}]{Figs/ProposedConcept}{
%        \psfrag{A}[c][c][0.75]{\shortstack{reflecting\\ \\transmission\\ \\line\\ \\array\\ \\(R-TLA)}}
%        \psfrag{B}[c][c][0.75]{\shortstack{calibration\\ \\transmission\\ \\line\\ \\array\\ \\(C-TLA)}}
%        \psfrag{C}[c][c][0.75]{Rotman lens}
%        \psfrag{D}[l][l][0.9]{$\omega=f(\alpha; \alpha_\text{in})$}
%        \psfrag{E}[c][c][0.75]{$\Psi_\text{in}(\omega)$}
%        \psfrag{F}[c][c][0.75]{$\Psi_\text{out}(\alpha,\omega)$}
%        \psfrag{G}[l][l][0.75]{$f$: spatial decomposition}
%        \psfrag{H}[l][l][0.75]{$\alpha$: flexible resolution}
%        \psfrag{I}[l][l][0.75]{$\alpha_\text{in}$: arbitrary input}
%        \psfrag{a}[c][c][0.75]{$\omega_1$}
%        \psfrag{b}[c][c][0.75]{$\omega_2$}
%        \psfrag{c}[c][c][0.75]{$\omega_3$}
%        \psfrag{d}[c][c][0.75]{$\omega_4$}
%        \psfrag{e}[c][c][0.75]{$\omega_5$}
%        \psfrag{f}[c][c][0.75]{$\omega_M$}
%        \psfrag{i}[c][c][0.75]{$\omega$}
%        \psfrag{j}[c][c][0.75]{$\Psi_\text{in}$}
%        \psfrag{h}[c][c][0.75]{$\alpha_\text{in}$}
%        \psfrag{g}[c][c][0.75]{$\alpha$}
%        }
        \caption{Proposed concept of flexible-resolution and arbitrary-input RL-SD.}
   \label{FIG:ProposedConcept}
\end{figure}

Thus, uniformizing or manipulating the resolution function, $\varrho(\omega)$, of the RL-SD, is highly desirable. We will show in Sec.~\ref{sec:synthesis} how exactly this can be achieved by adjusting the position of the output ports, as represented in~\figref{FIG:ProposedConcept}, since $\omega=f(\alpha)$.

Another desirable enhancement of the RL-SD would be to provide flexibility in the location of the input port, for the following two reasons. First, when the desired resolution function leads to a design with an output port at the same location as the input port, as for instance in \figref{FIG:FuscoRLSD}, a circulator or directional coupler is required to separate the input signal and output decomposed signal. Second, the resolution-enhanced design may have no output port and no room for an input port at the center of the structure. Such an enhancement can be realized by the insertion of a calibration transmission line array (C-TLA) between the lens and the R-TLA, as shown in~\figref{FIG:ProposedConcept} with $\omega=f(\alpha;\alpha_\text{in})$ and as will be detailed in Sec.~\ref{sec:C-TLA}.

\section{Synthesis}\label{sec:synthesis}

\subsection{Frequency-Position Relationship}\label{sec:freq_pos_rel}

The relationship existing between the spectral component $\omega$ and the position $\alpha$ on the left contour of the RL may be found by equating the function $\Delta\phi(\omega)$ corresponding to the R-TLA in Fig.~\ref{FIG:FuscoRLSD}, which is given by~\eqref{DeltaPhi1} and may be generalized to
\begin{equation}\label{EQ:DeltaPhiRTLA}
\Delta\phi(\omega)=2\pi N\left(\frac{\omega}{\omega_0}\right)-2\pi N,
\end{equation}
and the function $\Delta\phi(\omega)$ corresponding to the RL beamforming in Fig.~\ref{FIG:RotmanLens}, which is given, according to antenna array theory, by
\begin{equation}\label{EQ:DeltaPhiAlpha}
\Delta\phi(\omega)
=\frac{\omega}{c}d\sin{\psi}
=\frac{\omega}{c}d\gamma\sin{\alpha},
\end{equation}
where~\eqref{EQ:Gamma} has been used in the last equality.
Equating~\eqref{EQ:DeltaPhiRTLA} and~\eqref{EQ:DeltaPhiAlpha}, and solving the resulting equation for $\omega$ yields the sought after relationship:
\begin{equation}\label{EQ:Omega}
\omega(\alpha)=\frac{2\pi N\omega_0c}{2\pi Nc-\omega_0d\gamma\sin{\alpha}}.
\end{equation}
Assuming that the RL-SD output port angle $\alpha$ is limited to [$-\alpha_0,\alpha_0$], the maximal frequency and minimal frequency are found by replacing $\alpha$ in~\eqref{EQ:Omega} with $\alpha_0$ and $-\alpha_0$, respectively, which yields
\begin{subequations}\label{Max_Min}
\begin{equation}\label{Max}
 \omega_\text{max}=\frac{2\pi N\omega_0c}{2\pi Nc-\omega_0d\gamma\sin{\alpha_0}},
\end{equation}
\begin{equation}\label{Min}
    \omega_\text{min}=\frac{2\pi N\omega_0c}{2\pi Nc+\omega_0d\gamma\sin{\alpha_0}},
\end{equation}
\end{subequations}
respectively.

\begin{figure}[h!t]
    \centering
    \includegraphics[width=0.9\columnwidth]{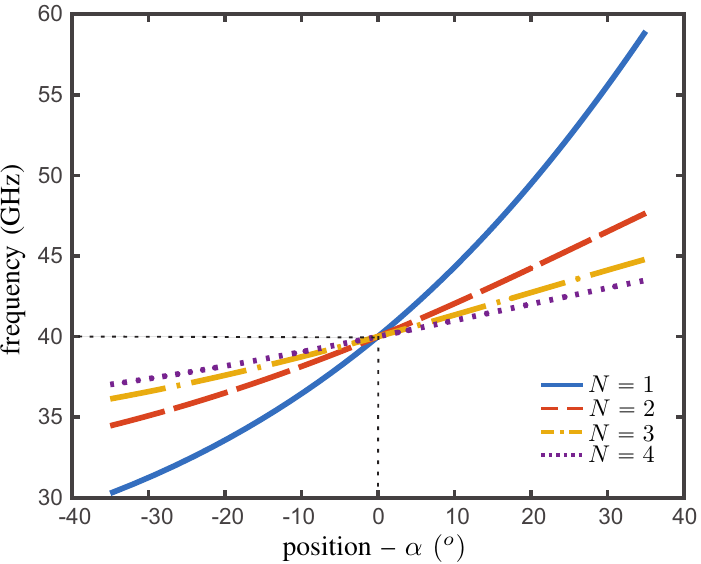}
       % \psfragfig[width=0.8\linewidth, trim={0in 0in 0in 0in}]{Figs/alpha_f}{
%        \psfrag{A}[c][c][0.8]{position -- $\alpha\ (^ o)$ }
%        \psfrag{B}[c][c][0.8]{frequency (GHz)}
%        \psfrag{a}[l][l][0.7]{$N=1$}
%        \psfrag{b}[l][l][0.7]{$N=2$}
%        \psfrag{c}[l][l][0.7]{$N=3$}
%        \psfrag{d}[l][l][0.7]{$N=4$}
%       }
        \caption{Decomposition of frequency versus position on the left contour of the RL-SD in~\figref{FIG:FuscoRLSD}, determined by the angle $\alpha$~(\figref{FIG:RotmanLens}), for different values of $N$ [Eq.~(\ref{DeltaL})], with $2\pi\omega_0=40$~GHz and $d=\lambda_0/2$.}
   \label{Fig:Omega}
\end{figure}

The function $\omega(\alpha)$ in~\eqref{EQ:Omega} is plotted in~\figref{Fig:Omega} for different values of $N$. It clearly appears that $\omega$ is a monotonically increasing function of $\alpha$ within the range [$-\alpha_0,\alpha_0$]. The figure also reveals that $N$ provides a parameter to trade off bandwidth and resolution, since the $\alpha$ variable will eventually be restricted to the discrete output ports. In~\figref{Fig:Omega}, the operating bandwidth decreases from 25.0~GHz to 5.8~GHz as $N$ increase from 1 to 4, while the resolution is enhanced by the same factor (on average in case of nonuniform port distribution).

\subsection{Calibration Waveguide Array for Arbitrary Input Position}\label{sec:C-TLA}

\begin{figure}[h!t]
    \centering
    \includegraphics[width=0.9\columnwidth]{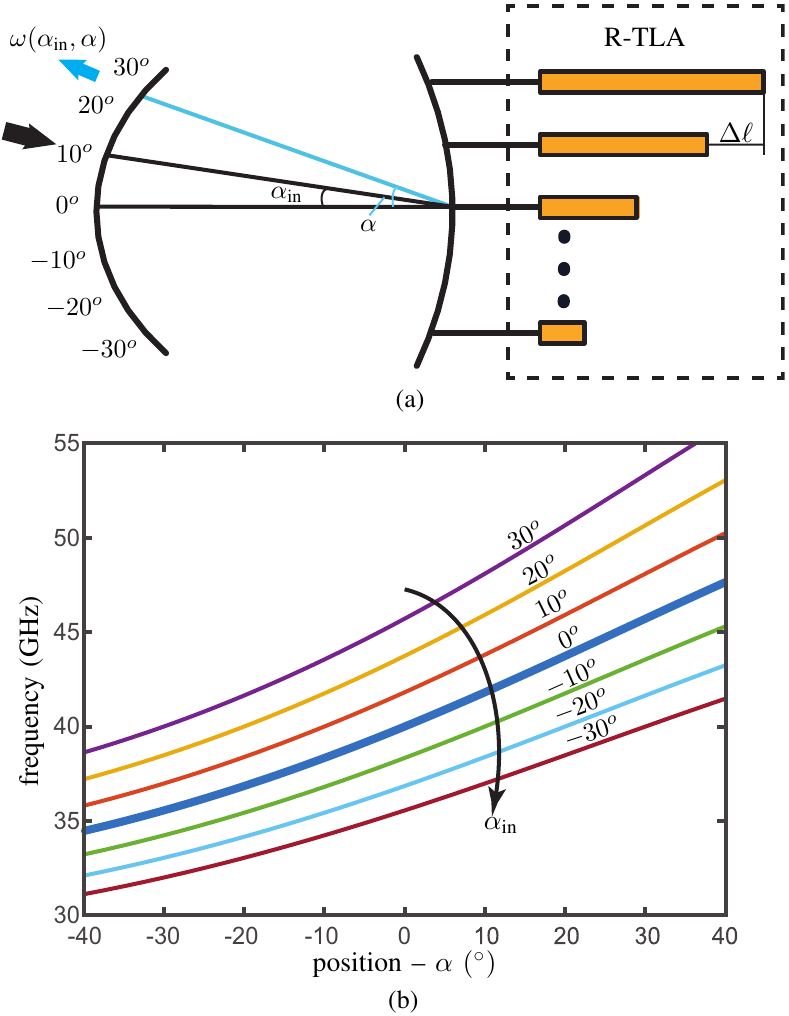}
        %\psfragfig[width=0.9\linewidth, trim={0in 0in 0in 0in}]{Figs/CalibrationDeviation}{
%        \psfrag{A}[c][c][0.8]{ position -- $\alpha\ (^\circ)$ }
%        \psfrag{B}[c][c][0.8]{frequency (GHz)}
%        \psfrag{p}[l][l][0.75]{$\alpha_\text{in}$}
%        \psfrag{a}[r][r][0.75]{$30^o$}
%        \psfrag{b}[r][r][0.75]{$20^o$}
%        \psfrag{c}[r][r][0.75]{$10^o$}
%        \psfrag{d}[r][r][0.75]{$0^o$}
%        \psfrag{e}[r][r][0.75]{$-10^o$}
%        \psfrag{f}[r][r][0.75]{$-20^o$}
%        \psfrag{g}[r][r][0.75]{$-30^o$}
%        \psfrag{h}[r][r][0.75]{$\alpha_\text{in}$}
%        \psfrag{i}[c][c][0.75]{$\alpha$}
%        \psfrag{w}[r][r][0.75]{$\omega(\alpha_\text{in},\alpha)$}
%        \psfrag{E}[c][c][0.75]{(a)}
%        \psfrag{F}[c][c][0.75]{(b)}
%        \psfrag{C}[c][c][0.8]{$\alpha_\text{in}$}
%        \psfrag{K}[c][c][0.8]{$\Delta\ell$}
%        \psfrag{R}[c][c][0.8]{R-TLA}
%        }
        \caption{Frequency deviation [Eq.~\eqref{EQ:OmegaDeviation}] caused by displacement of the input port from the middle ($\alpha_\text{in}=0$, thick curve) of the RL-SD structure.~(a)~Input port location $\alpha_\text{in}$ and decomposition function $f(\alpha,\alpha_\text{in})$.~(b)~Frequency versus position for $\alpha_\text{in}=0^o,\pm10^o,\pm20^o,\pm30^o$.}
   \label{FIG:Deviation}
\end{figure}

Changing the position of the input port in the RL-SD of \figref{FIG:FuscoRLSD}, as illustrated in \figref{FIG:Deviation}(a), would naturally modify the optical path lengths across the structure and hence alter its response. According to~\eqref{EQ:Gamma}, $\alpha_\text{in}=0$ (Fig.~\ref{FIG:FuscoRLSD}) corresponds to $\psi=0$. Therefore, since the sine function in~\eqref{EQ:Gamma} is monotonous in its range of interest about $0$, $\alpha_\text{in}\neq 0$ corresponds to $\psi=\psi'\neq 0$. Such nonzero angle, would correspond in the beamformer to the phase shift function $\Delta\phi_\text{in}=\frac{\omega}{c}\gamma d\sin{\alpha_\text{in}}$. This is a phase term that must now be added to~\eqref{EQ:DeltaPhiRTLA} in order to account for the effect of the displacement of the input port. Thus, Eq.~\eqref{EQ:DeltaPhiRTLA} generalizes to
\begin{equation}\label{EQ:DeltaPhiCal}
\Delta\phi(\omega)=2\pi N\left(\frac{\omega}{\omega_0}\right)-2\pi N-\frac{\omega}{c}\gamma d\sin{\alpha_\text{in}}.
\end{equation}

\begin{figure}[h!t]
    \centering
    \includegraphics[width=0.9\columnwidth]{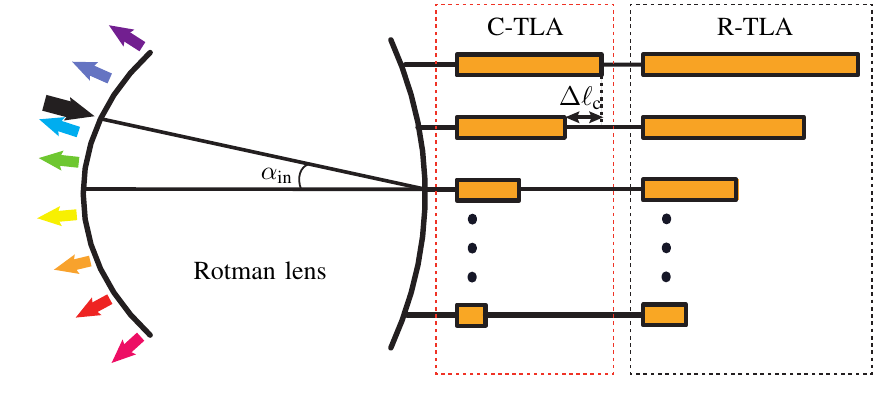}
        %\psfragfig[width=1\linewidth, trim={0in 0in 0in 0in}]{Figs/CorrectedDecomposer}{
%        \psfrag{C}[c][c][0.75]{C-TLA}
%        \psfrag{h}[c][c][0.75]{$\alpha_\text{in}$}
%        \psfrag{T}[c][c][0.75]{Rotman lens}
%        \psfrag{R}[c][c][0.75]{R-TLA}
%        \psfrag{l}[c][c][0.78]{$\Delta\ell_\text{c}$}
%        }
        \caption{Incorporation of calibration transmission line array (C-TLA) for suppressing the deviation effect in~\figref{FIG:Deviation}, i.e. merging all curves $\alpha_\text{in}\neq0$ to the curve $\alpha_\text{in}=0$ in that figure.}
   \label{FIG：Calibration}
\end{figure}

Equating~\eqref{EQ:DeltaPhiCal} to~\eqref{EQ:DeltaPhiAlpha}, and solving the resulting equation for $\omega$ yields the following generalization of~\eqref{EQ:Omega}:
\begin{equation}\label{EQ:OmegaDeviation}
\omega(\alpha)=\frac{2\pi N\omega_0c}{2\pi Nc-\omega_0d\gamma(\sin{\alpha}+\sin{\alpha_\text{in}})}.
\end{equation}
The frequency deviation computed by this relation is  plotted in~\figref{FIG:Deviation}(b). The graph also shows that the decomposition frequency range is shifted, towards higher and lower frequencies for positive and negative values of $\alpha_\text{in}$, respectively. In order to avoid this shift, we insert a \emph{calibration transmission line array (C-TLA)} between the lens and the R-TLA, as shown in \figref{FIG：Calibration}, to exactly compensate for the negative $\Delta\phi_\text{in}$ in~\eqref{EQ:DeltaPhiCal}. For this condition to be satisfied, the C-TLA must add the phase shift $+\frac{\omega}{c}\gamma d\sin{\alpha_\text{in}}$, and its lengths difference, $\Delta\ell_\text{c}$, must therefore satisfy the relation
\begin{equation}\label{EQ:CalEqual}
  2\beta_\text{e}(\omega_0)\Delta\ell_\text{c}
  =\frac{\omega}{c}\gamma d\sin{\alpha_\text{in}},
\end{equation}
which yields
\begin{equation}\label{EQ:Calibration}
  \Delta\ell_\text{c} = \frac{d}{2}\gamma\sqrt{\frac{1}{\epsilon_\text{e}}}\sin{\alpha_\text{in}}.
\end{equation}
After such calibration, the relevant formula for the frequency-position law is again~\eqref{EQ:Omega}.

\subsection{Frequency-Position Sampling for Flexible Resolution}

\begin{figure}[h!t]
    \centering
    \includegraphics[width=0.88\columnwidth]{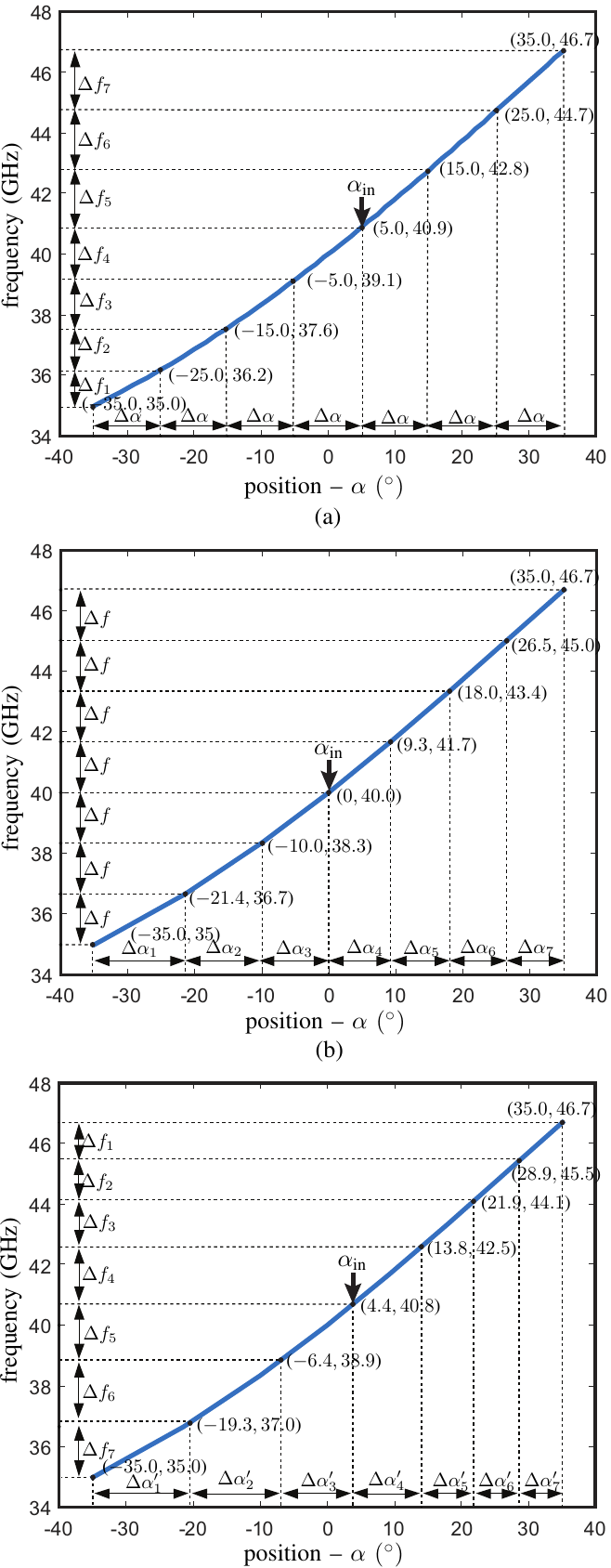}
        \caption{Resolution diversification of the RL-SD via modulation of the position distribution {$\alpha_m$}, $m=1,2,3,...,M$ for the following parameters: $2\pi\omega_0=40$~GHz, $d=\lambda_0/2$ and $N=2$. (a)~Uniform port distribution ($\Delta\alpha=\text{const}.$) leading to  nonuniform spectrum resolution ($\varrho \neq \text{const}.$)~[\figref{FIG:SpectrumAnalysis}(a)]. (b)~Nonuniform port distribution ($\Delta\alpha \neq \text{const}$.) leading to uniform spectrum resolution ($\varrho = \text{const}.$)~[\figref{FIG:SpectrumAnalysis}(b)]. (c) Nonuniform port distribution ($\Delta\alpha \neq \text{const}.$) leading to reversed spectrum resolution of uniform port distribution~[\figref{FIG:SpectrumAnalysis}(c)]. (Note that the curves in (a), (b) and (c) are exactly the same, only the $(\omega,\alpha)$ sampling and position of input port $(\alpha_\text{in})$ differs between them.)}
   \label{FIG:Omegaabc}
\end{figure}

Equation~\eqref{EQ:Omega} confirms and quantifies the dependence between the output port position and the frequency, that was qualitatively described in Sec.~\ref{sec:resol_in_port_divers}. Figure~\ref{FIG:Omegaabc} shows how this relation, which is independent of the input port location after calibration, can be manipulated to provide the spectrum analysis diversification illustrated in \figref{FIG:SpectrumAnalysis}.

\begin{figure}[h!t]
    \centering
    \includegraphics[width=0.85\columnwidth]{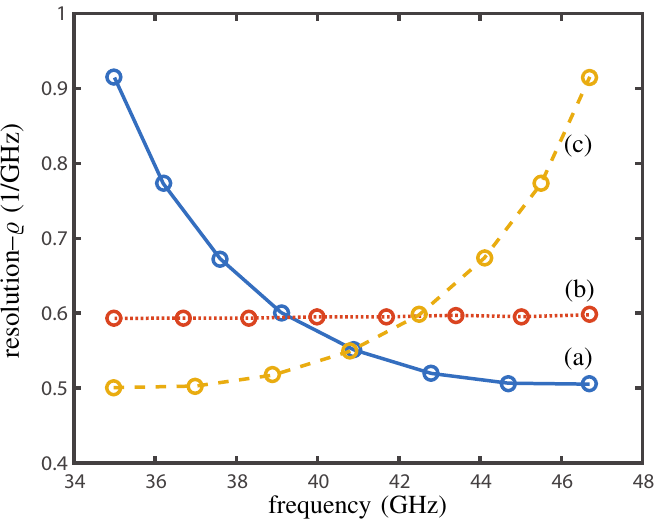}
        %\psfragfig[width=0.75\linewidth, trim={0in 0in 0in 0in}]{Figs/Resolution1}{
%        \psfrag{B}[c][c][0.75]{frequency (GHz)}
%        \psfrag{a}[c][c][0.75]{(c)}
%        \psfrag{b}[c][c][0.75]{(b)}
%        \psfrag{c}[c][c][0.75]{(a)}
%        \psfrag{A}[c][c][0.78]{resolution--$\varrho$ (1/GHz)}
%        }
        \caption{Resolution ($\varrho$) versus frequency ($\omega$) for the RL-SDs in~\figref{FIG:Omegaabc}.}
   \label{Fig:Resolution}
\end{figure}

Figure~\ref{FIG:Omegaabc}(a) shows the case of uniform angular port distribution ($\Delta\alpha=\text{const}$), leading to decreasing resolution with increasing frequency ($\partial\varrho/\partial\omega<0$) and corresponding to \figref{FIG:SpectrumAnalysis}(a). Figure~\ref{FIG:Omegaabc}(b) shows the case the nonuniform angular port distribution ($\Delta\alpha\neq\text{const}$) that leads to uniform resolution ($\partial\varrho/\partial\omega=0$), corresponding to \figref{FIG:SpectrumAnalysis}(b). Figure~\ref{FIG:Omegaabc}(c) shows a case of nonuniform angular port distribution leading to reversed resolution compared to \figref{FIG:Omegaabc}(a), ($\partial\varrho/\partial\omega>0$) and corresponding to \figref{FIG:SpectrumAnalysis}(c). The resolution versus frequency curves for the three cases are compared in~\figref{Fig:Resolution}, corresponding to the qualitative result of \figref{FIG:SpectrumAnalysis}(d).

\subsection{Range and Resolution Tuning by Input Switching}

Once the RL-SD has been built for the desired frequency range and resolution, according to the methodology outlined in the previous sections, we may be interested, for applications requiring adaptivity, to tune these parameters. Such a tunability functionality may be added by introducing port switching.

This may be seen following a procedure that is similar to that used in Sec.~\ref{sec:C-TLA}.
According to~\eqref{EQ:Gamma}, $\sin(\psi_\text{in})=\gamma\sin(\alpha_\text{in})$. Therefore, switching from the input located at $\alpha_\text{in}$ to the port located at $\alpha_\text{in}+\Delta\alpha_\text{in}$, leads to $\sin(\psi_\text{in}+\Delta\psi)=\gamma\sin(\alpha_\text{in}+\Delta\alpha_\text{in})$. These two cases correspond to the beamformer phase shift functions $\Delta\phi_\text{a}=\frac{\omega}{c}\gamma d\sin{\alpha_\text{in}}$ and $\Delta\phi_\text{b}=\frac{\omega}{c}\gamma d\sin{(\alpha_\text{in}+\Delta\alpha_\text{in})}$, respectively, which leads to the phase difference $\Delta\phi_\text{b}-\Delta\phi_\text{a}=\frac{\omega}{c}\gamma d[\sin{(\alpha_\text{in}+\Delta\alpha_\text{in})}-\sin{\alpha_\text{in}}]$, and hence to the new phase function
\begin{equation}\label{DeltaPhi6}
\Delta\phi(\omega)=2\pi N\left(\frac{\omega}{\omega_0}\right)-2\pi N-\frac{\omega}{c}\gamma d[\sin{(\alpha_\text{in}+\Delta\alpha_\text{in})}-\sin{\alpha_\text{in}}].
\end{equation}
Equating this relation to~\eqref{EQ:DeltaPhiAlpha} yields the tuning law
\begin{equation}\label{EQ:OmegaDivesity}
\omega(\alpha)=\frac{2\pi N\omega_0c}{2\pi Nc-\omega_0d\gamma(\sin{\alpha}+\sin{(\alpha_\text{in}+\Delta\alpha_\text{in})}-\sin{\alpha_\text{in}})}.
\end{equation}

\begin{figure}[h!t]
    \centering
    \includegraphics[width=0.85\columnwidth]{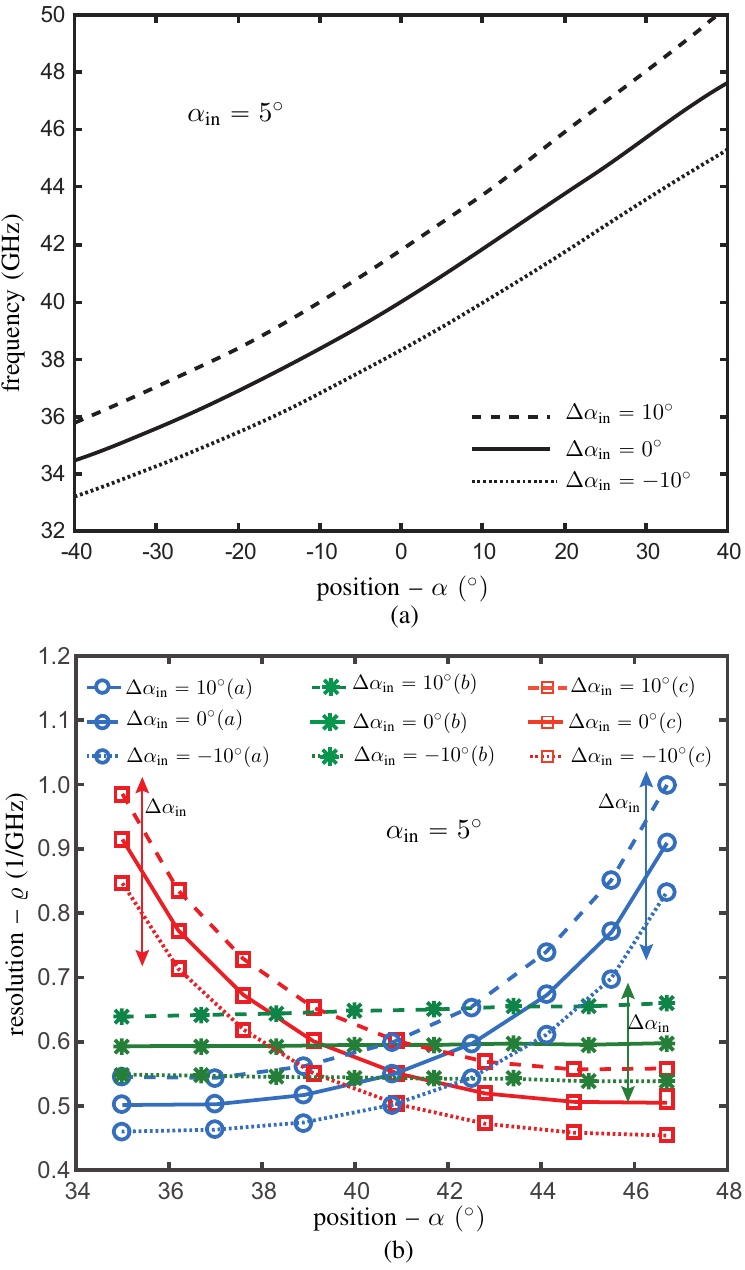}
        %\psfragfig[width=0.85\linewidth, trim={0in 0in 0in 0in}]{Figs/ResolutionDiversity2}{
%        \psfrag{A}[c][c][0.75]{position -- $\alpha\ (^\circ)$}
%        \psfrag{B}[c][c][0.75]{frequency (GHz)}
%        \psfrag{C}[c][c][0.75]{resolution -- $\varrho$ (1/GHz)}
%        \psfrag{b}[c][c][0.75]{(b)}
%        \psfrag{a}[c][c][0.75]{(a)}
%        \psfrag{x}[l][l][0.65]{$\Delta\alpha_\text{in}=10^\circ$}
%        \psfrag{y}[l][l][0.65]{$\Delta\alpha_\text{in}=0^\circ $}
%        \psfrag{z}[l][l][0.65]{$\Delta\alpha_\text{in}=-10^\circ$}
%        \psfrag{D}[c][c][0.75]{resolution $\varrho$ (1/GHz)}
%        \psfrag{c}[l][l][0.6]{$\Delta\alpha_\text{in}=10^\circ (a)$}
%        \psfrag{d}[l][l][0.6]{$\Delta\alpha_\text{in}=0^\circ (a)$}
%        \psfrag{e}[l][l][0.6]{$\Delta\alpha_\text{in}=-10^\circ (a)$}
%        \psfrag{f}[l][l][0.6]{$\Delta\alpha_\text{in}=10^\circ (b)$}
%        \psfrag{g}[l][l][0.6]{$\Delta\alpha_\text{in}=0^\circ (b)$}
%        \psfrag{h}[l][l][0.6]{$\Delta\alpha_\text{in}=-10^\circ (b)$}
%        \psfrag{i}[l][l][0.6]{$\Delta\alpha_\text{in}=10^\circ (c)$}
%        \psfrag{j}[l][l][0.6]{$\Delta\alpha_\text{in}=0^\circ (c)$}
%        \psfrag{k}[l][l][0.6]{$\Delta\alpha_\text{in}=-10^\circ (c)$}
%        \psfrag{T}[l][l][0.8]{$\alpha_\text{in}=5^\circ$}
%        \psfrag{D}[c][c][0.6]{$\Delta\alpha_\text{in}$}
%        }
        \caption{Tuning obtained by switching the input port (or vary $\Delta\alpha_\text{in}$) from the starting $\alpha_\text{in}=5^\circ$ using~\eqref{EQ:OmegaDivesity}. (a)~Frequency range tuning achieved with $\Delta\alpha_\text{in} = -10^\circ, 0^\circ, 10^\circ$. (b)~Application of the three sampling strategies shown in~\figref{FIG:Omegaabc} for each of the three port-switched curve in (a), showing also frequency resolution tuning.}
   \label{Fig:DiversInput}
\end{figure}

Figure~\ref{Fig:DiversInput}(a) plots this $\omega(\alpha)$ relation, after calibration, under input port switching tuning from $\alpha_\text{in}=5^\circ$. The curve $\Delta\alpha_\text{in}=0^\circ$ in this figure corresponds to the (unique) curve in Fig.~\ref{FIG:Omegaabc}, while the curves $\Delta\alpha_\text{in}=\pm 10^\circ$ represent shifted curves resulting from switching from $\alpha_\text{in}=5^\circ$ to $\alpha_\text{in}+\Delta\alpha_\text{in}=15^\circ$ and $\alpha_\text{in}+\Delta\alpha_\text{in}=-5^\circ$. As may be seen, port switching results in frequency \emph{range} tuning ([$f_\text{min},f_\text{max}$]). Figure~\ref{Fig:DiversInput}(b) shows the port switching tuning achieved for different sampling strategies, and indicates that port switching also leads to frequency \emph{resolution} thing ([$\varrho(\alpha)$]). Note that the frequency range and resolution tunings are not independent from each other: the resolution variation follows the same trend as the range.

\section{Design Procedure}\label{sec:Procedure}

The overall design procedure of the flexible-resolution, arbitrary-input and tunable RL-SD is as follow:

\begin{enumerate}
  \item Specify the operating frequency range [$f_\text{min},f_\text{max}$].
  \item Choose the focal lengths $f_1$ and $f_2$ based on trade-off between device size and number of port (and hence resolution).
  \item Select a set of parameters $\omega_0$, $N$, $d$, $\alpha_0$ $\gamma$ to accommodate the [$f_\text{min},f_\text{max}$] according to~\eqref{Max_Min}.
  \item Determine the RL geometry in terms of its transmission line lengths, left contour and right contour according to~\eqref{Eq:TL},~\eqref{Eq:ContourL} and~\eqref{Eq:ContourR}, respectively.
  \item Determine a number of ports based on the room available and their position on the left contour using~\eqref{EQ:Omega}.
  \item Calculate the transmission line width for $50\ \Omega$ characteristic impedance and connect all the output points to the output ports with these lines using a taper for broadband impedance matching.
  \item Calculate transmission line length differences $\Delta\ell$ of the R-TLA according to~\eqref{DeltaL} and $\Delta\ell_\text{c}$ of the C-TLA accoding to~\eqref{EQ:Calibration}.
  \item Interconnect the C-TLA, R-TLA and RL.
\end{enumerate}

\section{Results}\label{sec:Results}

In order to demonstrate the concepts developed in the previous sections, we present here the design, simulation, fabrication and measurement of two RL-SD prototypes, corresponding to the frequency-position sampling strategies to~\figref{FIG:Omegaabc}(a) and~\figref{FIG:Omegaabc}(b), respectively.

\begin{figure}[h!t]
    \centering
    \includegraphics[width=0.8\columnwidth]{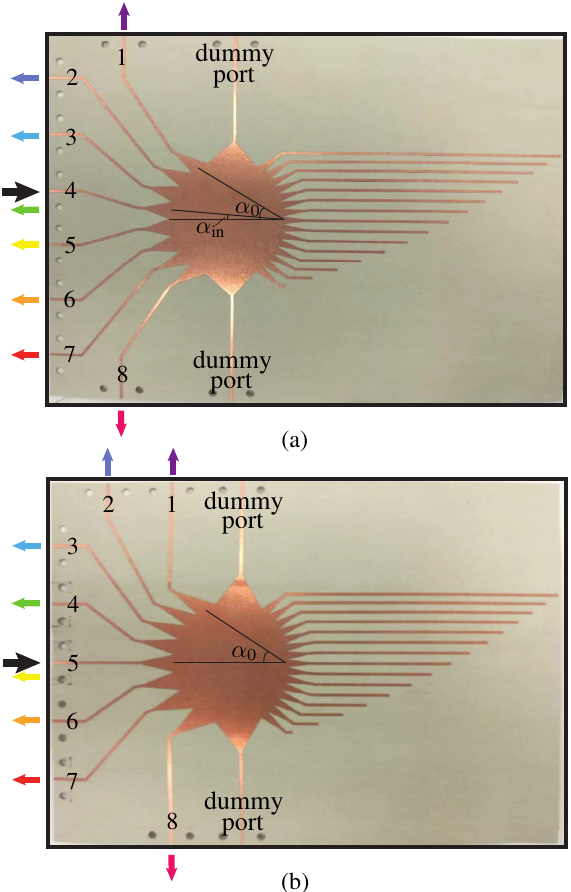}
        %\psfragfig[width=0.65\linewidth, trim={0in 0in 0in 0in}]{Figs/photos}{
%        \psfrag{D}[c][c][0.75]{dummy}
%        \psfrag{E}[c][c][0.75]{port}
%        \psfrag{a}[c][c][0.7]{1}
%        \psfrag{b}[c][c][0.7]{2}
%        \psfrag{c}[c][c][0.7]{3}
%        \psfrag{d}[c][c][0.7]{4}
%        \psfrag{e}[c][c][0.7]{5}
%        \psfrag{f}[c][c][0.7]{6}
%        \psfrag{g}[c][c][0.7]{7}
%        \psfrag{h}[c][c][0.7]{8}
%        \psfrag{i}[c][c][0.7]{$\alpha_0$}
%        \psfrag{j}[c][c][0.7]{$\alpha_\text{in}$}
%        \psfrag{x}[c][c][0.7]{(b)}
%        \psfrag{y}[c][c][0.7]{(a)}
%        }
        \caption{RL-SD prototypes. (a)~Uniform port distribution with off-axis input~[\figref{FIG:Omegaabc}~(a)]. (b)~Uniform spectrum resolution~[\figref{FIG:Omegaabc}~(b)]. }
   \label{Fig:Prototype}
\end{figure}

Figure~\ref{Fig:Prototype} shows the photographs of the two prototypes, with~\figref{Fig:Prototype}(a) and~\figref{Fig:Prototype}(a) corresponding to~\figref{FIG:Omegaabc}(a) and~\figref{FIG:Omegaabc}(b), respectively. The two designs have the following parameters in common. They are
fabricated on a 10-mil thick Rogers 6002 dielectric substrate with dielectric constant 2.94 and loss tangent 0.0012, for the frequency range [35, 46.7]~GHz and $f_0=40$ GHz. They include 8 output ports and 15 RL transmission lines with corresponding C-TLA and R-TLA. The two designs also share the same RL symmetry, with $f_1=28.4$ mm, $f_2=25.6$ mm, $\alpha_0=\psi_0=35^\circ$ and $d=\lambda_0/2$. Their R-TLA transmission line length difference is $\Delta\ell= 4.77$ mm with $N=2$ according to~\eqref{Eq:TL}.

The first prototype [\figref{Fig:Prototype}(a)] has uniformly distributed output ports with port~4 as the input port at $\alpha_\text{in}=5^\circ$ and C-TLA lines length difference $\Delta\ell_\text{c}=0.066$ mm according to~\eqref{EQ:Calibration}. The second prototype [\figref{Fig:Prototype}(b)] has nonuniformly distributed ports for uniform resolution with port~5 as the input port with $\alpha_\text{in}=0^\circ$ and C-TLA lines length difference $\Delta\ell_\text{c}=0$.

\begin{figure*}
  \centering
  \includegraphics[width=0.88\linewidth, trim={0in 0in 0in 0in}]{{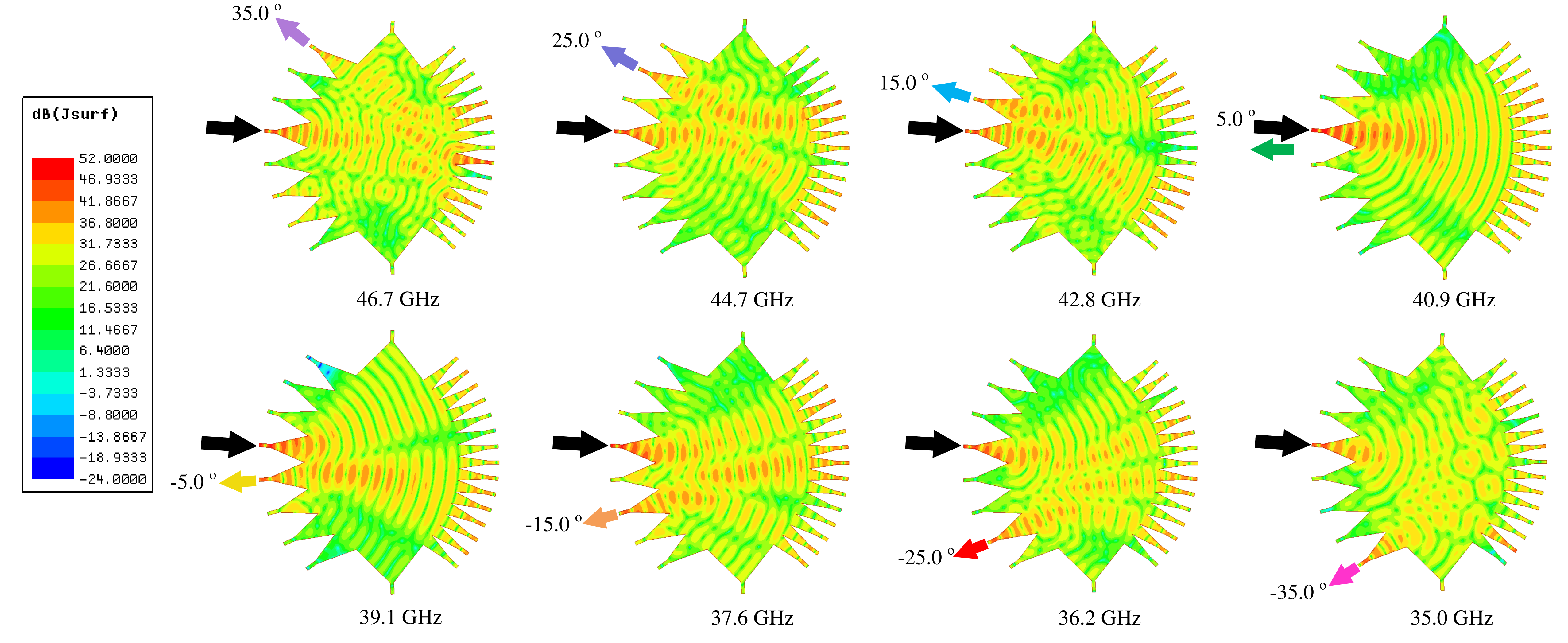}}
  \caption{Current distribution in RL-SD with uniform port distribution, nonuniform spectrum resolution and $\alpha_\text{in}=5^ o $ ~[\figref{FIG:Omegaabc} (a)].}\label{Fig:Current 1}
\end{figure*}

\begin{figure*}
  \centering
  \includegraphics[width=0.88\linewidth, trim={0in 0in 0in 0in}]{{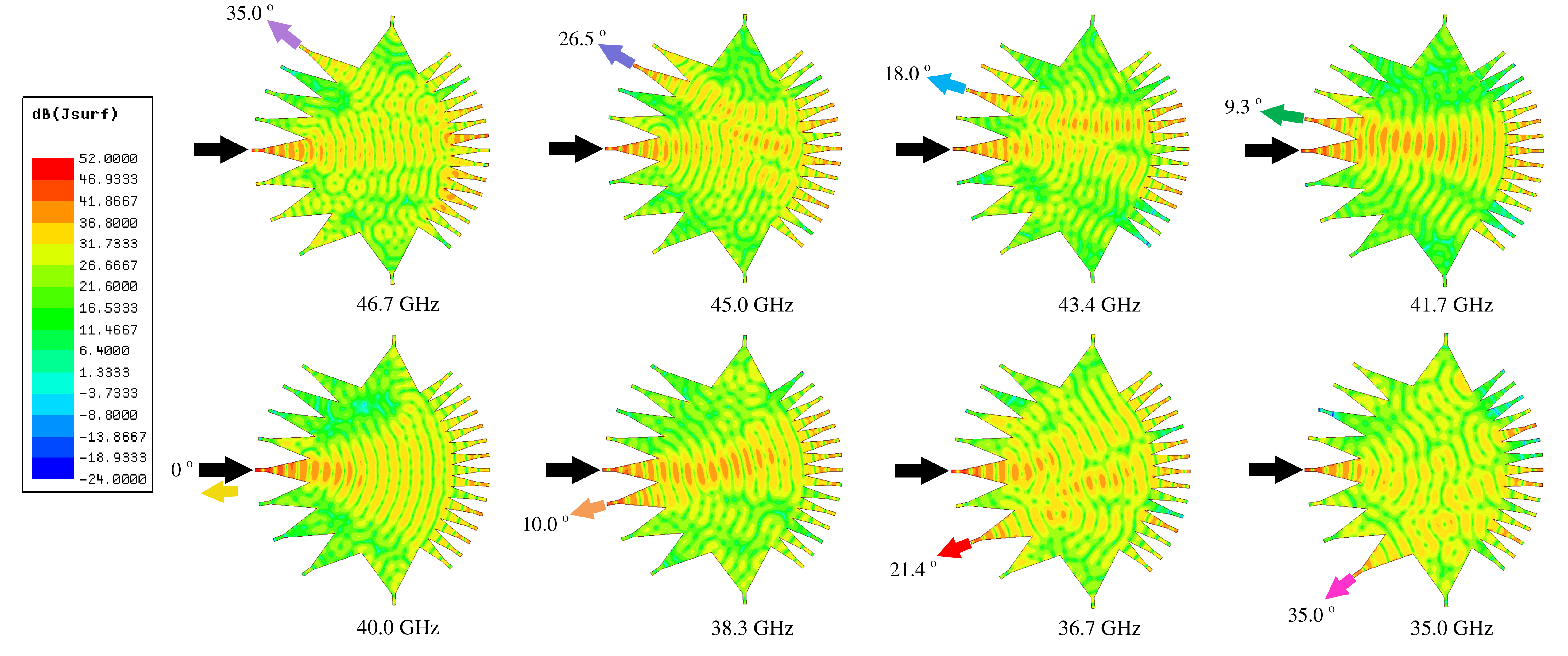}}
  \caption{Current distribution in RL-SD with nonuniform port distribution, uniform spectrum resolution and $\alpha_\text{in}=0^ o$ ~[\figref{FIG:Omegaabc} (b)].}\label{Fig:Current 2}
\end{figure*}

The RL-SDs are simulated with the help of full-wave simulation solver ANSYS Electronics~2015. Figures~\ref{Fig:Current 1} and~\ref{Fig:Current 2} show the simulated current distributions in the RL part of the RL-SD of the two prototypes. They clearly show that the currents flow to the expected output ports, corresponding to the designs in~\figref{FIG:Omegaabc}(a) and~\figref{FIG:Omegaabc}(b).

\begin{figure}[h!t]
    \centering
    \includegraphics[width=0.9\columnwidth]{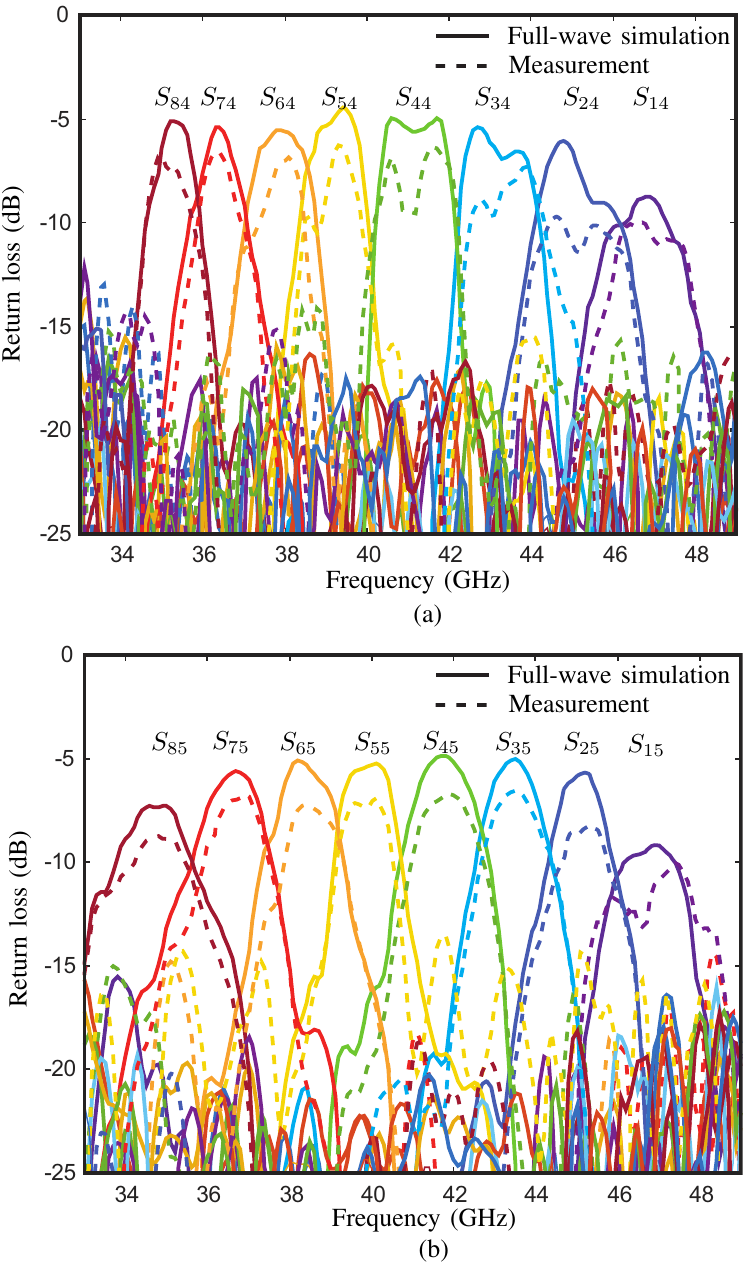}
        %\psfragfig[width=0.85\linewidth, trim={0in 0in 0in 0in}]{Figs/SP}{
%        \psfrag{A}[c][c][0.75]{Frequency (GHz)}
%        \psfrag{B}[c][c][0.75]{Return loss (dB)}
%        \psfrag{C}[l][l][0.75]{Full-wave simulation}
%        \psfrag{D}[l][l][0.75]{Measurement}
%        \psfrag{a}[l][l][0.75]{(a)}
%        \psfrag{b}[l][l][0.75]{(b)}
%        \psfrag{E}[c][c][0.75]{$S_{84}$}
%        \psfrag{F}[c][c][0.75]{$S_{74}$}
%        \psfrag{G}[c][c][0.75]{$S_{64}$}
%        \psfrag{H}[c][c][0.75]{$S_{54}$}
%        \psfrag{I}[c][c][0.75]{$S_{44}$}
%        \psfrag{J}[c][c][0.75]{$S_{34}$}
%        \psfrag{K}[c][c][0.75]{$S_{24}$}
%        \psfrag{L}[c][c][0.75]{$S_{14}$}
%        \psfrag{M}[c][c][0.75]{$S_{45}$}
%        \psfrag{N}[c][c][0.75]{$S_{35}$}
%        \psfrag{O}[c][c][0.75]{$S_{25}$}
%        \psfrag{P}[c][c][0.75]{$S_{15}$}
%        \psfrag{Q}[c][c][0.75]{$S_{85}$}
%        \psfrag{R}[c][c][0.75]{$S_{75}$}
%        \psfrag{S}[c][c][0.75]{$S_{65}$}
%        \psfrag{T}[c][c][0.75]{$S_{55}$}
%        }
        \caption{Comparison of full-wave simulated and measured scattering parameters. (a)~Uniform port distribution with off-axis input~[\figref{FIG:Omegaabc}~(a)]. (b)~Uniform spectrum resolution~[\figref{FIG:Omegaabc}~(b)]. Each color corresponds to the spectrum reaching an port and should not be confused with the port colors in \figref{Fig:Prototype} that represent only the \emph{maximal} spectral energy.}
   \label{Fig:SP}
\end{figure}

Finally, \figref{Fig:SP} plots the simulated and measured scatting parameters for the designs. The main beams are centered at the expected frequency with an isolation in the order of 10~dB to the parastic side-lobes, corresponding to leakage to other ports, due to port coupling. The measured results feature about 2~dB more loss than the simulation results, due to fabrication imperfection and impedance mismatched\footnote{The idle output ports were imperfectly terminated by flexible absorbing material.}. The measured results also exhibit lower isolation. This is especially the case at the output ports that also used as input ports (port~4 in~\figref{Fig:SP}(a) and port 5~in \figref{Fig:SP}(b)) and this may be understood as follows:

All the frequencies, forming the spectrum of the signal to analyze, are injected into the RL-SD at the input port, get directed to the design upon first reflection by the R-TLA and, due to imperfect matching, they are then reflected back for a second \emph{inverse-direction} round trip. Due to reciprocity, all the frequencies are scattered back to the input port at this second round trip, and leak into that port, which leads to $S_{\text{in-port},\text{in-port}}(\omega\neq\omega_\text{intended at in-port})$ corresponding to the highest parasitic lobes.


\section{Conclusion}

An enhanced design -- in terms of resolution flexibility, input port position arbitrariness and frequency-range tunability -- of the planar Rotman lens spectrum decomposer (RL-SD) has been presented and demonstrated both theoretically and experimentally. Given these additional features, adding upon the  planar, low-cost, integrable and frequency-scalable features, the RL-SD may find wide applications across the entire electromagnetic spectrum from microwaves to optics.

\bibliographystyle{IEEEtran}
\bibliography{IEEEabrv,Xiaoyi_Reference}

\begin{thebibliography}{10}
\providecommand{\url}[1]{#1}
\csname url@samestyle\endcsname
\providecommand{\newblock}{\relax}
\providecommand{\bibinfo}[2]{#2}
\providecommand{\BIBentrySTDinterwordspacing}{\spaceskip=0pt\relax}
\providecommand{\BIBentryALTinterwordstretchfactor}{4}
\providecommand{\BIBentryALTinterwordspacing}{\spaceskip=\fontdimen2\font plus
\BIBentryALTinterwordstretchfactor\fontdimen3\font minus
  \fontdimen4\font\relax}
\providecommand{\BIBforeignlanguage}[2]{{%
\expandafter\ifx\csname l@#1\endcsname\relax
\typeout{** WARNING: IEEEtran.bst: No hyphenation pattern has been}%
\typeout{** loaded for the language `#1'. Using the pattern for}%
\typeout{** the default language instead.}%
\else
\language=\csname l@#1\endcsname
\fi
#2}}
\providecommand{\BIBdecl}{\relax}
\BIBdecl

\bibitem{Book:Waldman_2002_LightIntro}
G.~Waldman, \emph{Introduction to light: The physics of light, vision, and
  color}.\hskip 1em plus 0.5em minus 0.4em\relax Courier Corporation, 2002.

\bibitem{Book:Knittl_1976_ThinFilms}
Z.~Knittl, \emph{Optics of thin films: an optical multilayer theory}.\hskip 1em
  plus 0.5em minus 0.4em\relax Wiley London, 1976.

\bibitem{Book:Born_2013_PrinciplesofOptics}
M.~Born and E.~Wolf, \emph{Principles of optics: electromagnetic theory of
  propagation, interference and diffraction of light}.\hskip 1em plus 0.5em
  minus 0.4em\relax Elsevier, 2013.

\bibitem{Book:Saleh_2007_FundamentalsPhotonics}
B.~E.~A. Saleh and M.~C. Teich, ``Fundamentals of photonics,'' \emph{Wiley},
  2007.

\bibitem{Jour:Kok_1971_Colorimeter}
C.~J. Kok and M.~C. Boshoff, ``New spectrophotometer and tristimulus mask
  colorimeter,'' \emph{Appl. Opt.}, vol.~10, no.~12, pp. 2617--2620, 1971.

\bibitem{JOUR:2009_Gupta_TMTT_RTSA}
S.~Gupta, S.~Abielmona, and C.~Caloz, ``Microwave analog real-time spectrum
  analyzer ({RTSA}) based on the spectral-spatial decomposition property of
  leaky-wave structures,'' \emph{IEEE Trans. Microw. Theory Tech.}, vol.~57,
  no.~12, pp. 2989--2999, Dec 2009.

\bibitem{JOUR:Hard_1970_laser}
T.~M. Hard, ``Laser wavelength selection and output coupling by a grating,''
  \emph{Appl. Opt.}, vol.~9, no.~8, pp. 1825--1830, 1970.

\bibitem{JOUR:Hansch_1972_Laser_tuning}
T.~W. H{\"a}nsch, ``Repetitively pulsed tunable dye laser for high resolution
  spectroscopy,'' \emph{Appl. Opt.}, vol.~11, no.~4, pp. 895--898, 1972.

\bibitem{Jour:White_1947_Filter_Gratings}
J.~U. White, ``Gratings as broad band filters for the infra-red,'' \emph{J.
  Opt. Soc. Am.}, vol.~37, no.~9, pp. 713--717, 1947.

\bibitem{Jour:Knop_1978_Color_Filtering}
K.~Knop, ``Diffraction gratings for color filtering in the zero diffraction
  order,'' \emph{Appl. Opt.}, vol.~17, no.~22, pp. 3598--3603, 1978.

\bibitem{JOUR:Brackett_1990_WDM}
C.~A. Brackett, ``Dense wavelength division multiplexing networks: principles
  and applications,'' \emph{IEEE J. Sel. Areas Commun.}, vol.~8, no.~6, pp.
  948--964, Aug 1990.

\bibitem{JOUR:1978_Henry_TMTT_Grating_Multiplexer}
P.~S. Henry and J.~T. Ruscio, ``A low-loss diffraction grating frequency
  multiplexer,'' \emph{IEEE Trans. Microw. Theory Tech.}, vol.~26, no.~6, pp.
  428--433, Jun 1978.

\bibitem{Jour:Zmuda_1997_beamformer}
H.~Zmuda, R.~A. Soref, P.~Payson, S.~Johns, and E.~N. Toughlian, ``Photonic
  beamformer for phased array antennas using a fiber grating prism,''
  \emph{IEEE Photon. Technol.}, vol.~9, no.~2, pp. 241--243, Feb 1997.

\bibitem{JOUR:2009_Momeni_TMTT_Electrical_Prism}
O.~Momeni and E.~Afshari, ``Electrical prism: A high quality factor filter for
  millimeter-wave and terahertz frequencies,'' \emph{IEEE Trans. Microw. Theory
  Tech.}, vol.~57, no.~11, pp. 2790--2799, Nov 2009.

\bibitem{JOUR:2008_Afshari_TCS_2D_LC_Lattice}
E.~Afshari, H.~S. Bhat, and A.~Hajimiri, ``Ultrafast analog fourier transform
  using 2-{D} {LC} lattice,'' \emph{IEEE Trans. Circuits Syst}, vol.~55, no.~8,
  pp. 2332--2343, Sept 2008.

\bibitem{JOUR:2015_Gomez-Tornero_TMTT_SIW_Multiplexer}
J.~L. Gomez-Tornero, A.~J. Martinez-Ros, S.~Mercader-Pellicer, and
  G.~Goussetis, ``Simple broadband quasi-optical spatial multiplexer in
  substrate integrated technology,'' \emph{IEEE Trans. Microw. Theory Tech.},
  vol.~63, no.~5, pp. 1609--1620, May 2015.

\bibitem{Conf:Fusco_Multiplerxer_2012}
Y.~Zhang and V.~Fusco, ``Rotman lens used as a demultiplexer/multiplexer,'' in
  \emph{2012 42nd European Microwave Conference}, Oct. 2012, pp. 164--167.

\bibitem{Jour:Rotman_ProcIEEE_TrueTimeDelay}
R.~Rotman, M.~Tur, and L.~Yaron, ``True time delay in phased arrays,''
  \emph{Proc. IEEE}, vol. 104, no.~3, pp. 504--518, March 2016.

\bibitem{JOUR:Rotman_Rotmanlens_1963}
W.~Rotman and R.~Turner, ``Wide-angle microwave lens for line source
  applications,'' \emph{IEEE Trans. Antennas Propag.}, vol.~11, no.~6, pp.
  623--632, Nov. 1963.

\bibitem{JOUR:Hansen_Rotmanlenses_1991}
R.~C. Hansen, ``Design trades for {R}otman lenses,'' \emph{IEEE Trans. Antennas
  Propag.}, vol.~39, no.~4, pp. 464--472, Apr. 1991.

\bibitem{JOUR:2014_CJE_Vashist_ReviewRotamLens}
S.~Vashist, M.~K. Soni, and P.~K. Singhal, ``A review on the development of
  rotman lens antenna,'' \emph{Chinese Journal of Engineering}, 2014.

\bibitem{Thesis:Dong_RotmanLens_2009}
J.~Dong, ``Microwave lens designs: Optimization, fast simulation algorithms,
  and 360-degree scanning techniques,'' Ph.D. dissertation, Virginia Tech,
  2009.

\end{thebibliography}

\appendix

\subsection{RL Synthesis (\figref{FIG:RotmanLens})}\label{sec:RL_Synth}
The solution of~\eqref{EQ:RLBasics} provides the right contour and transmission line lengths of the RL. The transmission lines lengths, normalized as $\overline{w}=(w-w_0)/f_1$, are found as~\cite{JOUR:Rotman_Rotmanlens_1963}
\begin{equation}\label{Eq:TL}
\overline{w}(y_\text{a})=\frac{\sqrt{\epsilon_\text{e}}}{\sqrt{\epsilon_\text{r}}}\frac{-b\pm\sqrt{b^2-4ac}}{2a},\\
\end{equation}
where
\begin{subequations}
\begin{equation}
a=1-\left(\frac{1-\beta}{1-\beta\cos{\alpha_0}}\right)^2-\frac{1}{\epsilon_\text{r}}\left(\frac{\zeta}{\beta}\right)^2,
\end{equation}
\begin{equation}
b=-2+2\frac{\zeta^2}{\beta}+\frac{2(1-\beta)}{1-\beta\cos{\alpha_0}}-\frac{\zeta^4\sin^4{\alpha_0}(1-\beta)}{\epsilon_\text{r}(1-\beta\cos{\alpha_0})^2},\\
\end{equation}
\begin{equation}
c=\left(-\zeta^2+\frac{\zeta^2\sin^2{\alpha_0}}{1-\beta\cos{\alpha_0}}-\frac{\zeta^2\sin^2{\alpha_0}}{4\epsilon_\text{r}(1-\beta\cos{\alpha_0)}}\right)\frac{1}{\epsilon_\text{r}},
\end{equation}
\end{subequations}
with $\beta=f_2/f_1$ and $\zeta=\gamma y_\text{a}/f_1$. The right contour of the RL is found as
\begin{subequations}\label{Eq:ContourR}
\begin{equation}
    \label{ArrayContourX}
    x_R(y_\text{a})=\frac{\zeta^2\sin^2{\alpha_0}}{2\epsilon_\text{r}(\beta\cos{\alpha_0-1})}+ \frac{\sqrt{\epsilon_\text{e}}}{\sqrt{\epsilon_\text{r}}}\frac{(1-\beta)w}{\beta\cos{\alpha_0}-1},
\end{equation}
\begin{equation}
    \label{ArrayContourY}
    y_R(y_\text{a})=\zeta\left(\frac{1}{\sqrt{\epsilon_\text{r}}}-\frac{\sqrt{\epsilon_\text{e}}}{\epsilon_\text{r}}\frac{w}{\beta}\right).
\end{equation}
\end{subequations}

The left contour of the lens can be freely chosen. We opted for the circular arc
\begin{subequations}\label{Eq:ContourL}
\begin{equation}
    \label{BeamContourX}
    x_L[\theta(\alpha)]=\rho_0[1-\cos{\theta(\alpha)}],
\end{equation}
\begin{equation}
    \label{BeamContourY}
    y_L[\theta(\alpha)]=\rho_0\sin{\theta(\alpha)},
\end{equation}
\end{subequations}
where
\begin{subequations}
\begin{equation}
  \label{BeamCoutourParamRho}
  \rho_0=1-\frac{1-\beta^2}{2(1-\beta\cos{\alpha_0})},
\end{equation}
\begin{equation}
  \label{BeamCoutourParamPhi}
  \theta(\alpha)=\arcsin\left(\frac{1-\rho_0}{\rho_0}\sin{\alpha}\right)+\alpha.
\end{equation}
\end{subequations}
This last equation has been obtained by applying the law of cosines to the fixed triangle $0OF_2$, i.e.
\begin{subequations}
\begin{equation}
    \label{Law of Cosines}
    2(1-\rho_0)\beta\cos{\alpha_0}=(1-\rho_0)^2+\beta^2-\rho_0^2,
\end{equation}
and the law of sines for the varying triangle $0OF$, i.e.
\begin{equation}
    \label{Sines}
    \rho_0\sin{(\theta-\alpha)}=(1-\rho_0)\sin{\alpha},
\end{equation}
\end{subequations}
and solving this system for $\rho_0$ and $\phi$.

\end{document}